\documentclass[fleqn,usenatbib]{mnras}

\usepackage{newtxtext,newtxmath}

\usepackage[T1]{fontenc}
\DeclareRobustCommand{\VAN}[3]{#2}
\let\VANthebibliography\thebibliography
\def\thebibliography{\DeclareRobustCommand{\VAN}[3]{##3}\VANthebibliography}

\usepackage{graphicx}	
\usepackage{amsmath}	
\usepackage[caption=false]{subfig}

\newcommand{\ie}{{\it i.e.} }

\newcommand{\lcdm}{$\Lambda$CDM }
\newcommand{\lcdmnospace}{$\Lambda$CDM}
\newcommand{\be}{\begin{equation}}
\newcommand{\ee}{\end{equation}}

\title[The Deceleration of a Tilted Universe]{Observational constraints on the deceleration parameter in a tilted universe}

\author[K. Asvesta et al.]{
Kerkyra Asvesta,$^{1}$\thanks{E-mail: keasvest@auth.gr}
Lavrentios Kazantzidis,$^{2}$\thanks{E-mail: l.kazantzidis@uoi.gr}
Leandros Perivolaropoulos$^{2}$\thanks{E-mail: leandros@uoi.gr}
and Christos G. Tsagas$^{1,3}$\thanks{E-mail: tsagas@astro.auth.gr}
\\
$^{1}$Section of Astrophysics, Astronomy and Mechanics, Department of Physics, Aristotle University of Thessaloniki 54124, Greece\\
$^{2}$Department of Physics, University of Ioannina, GR-45110, Ioannina, Greece \\
$^{3}$Clare Hall, University of Cambridge, Herschel Road, Cambridge CB3 9AL, United Kingdom
}

\date{Accepted 2022 March 31. Received 2022 March 28; in original form 2022 February 14}

\pubyear{2022}

\begin{document}
\label{firstpage}
\pagerange{\pageref{firstpage}--\pageref{lastpage}}
\maketitle

\begin{abstract}
We study a parametrization of the deceleration parameter in a tilted universe, namely a cosmological model equipped with two families of observers. The first family follows the smooth Hubble flow, while the second is the real observers residing in a typical galaxy inside a bulk flow and moving relative to the smooth Hubble expansion with finite peculiar velocity. We use the compilation of Type
Ia Supernovae (SnIa) data, as described in the Pantheon dataset, to find the quality of fit to the data and study the redshift evolution of the deceleration parameter. In so doing, we consider two alternative scenarios, assuming that the bulk-flow observers live in the $\Lambda$CDM and in the Einstein-de Sitter universe. We show that a tilted Einstein-de Sitter model can reproduce the recent acceleration history of the universe, without the need of a cosmological constant or dark energy, by simply taking into account linear effects of peculiar motions. By means of a Markov Chain Monte Carlo (MCMC) method, we also constrain the magnitude and the uncertainties of the parameters of the two models. From our statistical analysis, we find that the tilted Einstein-de Sitter model, equipped with one or two additional parameters that describe the assumed large-scale velocity flows, performs similar to the standard $\Lambda$CDM paradigm in the context of model selection criteria (Akaike Information Criterion and Bayesian Information Criterion). 
\end{abstract}



\section{Introduction}
\label{sec:Introduction}

The discovery of the accelerated expansion of the universe \cite{SupernovaSearchTeam:1998fmf,SupernovaCosmologyProject:1998vns} has restored the idea of the cosmological constant in the last 20 years \cite{Carroll:2000fy}, rendering the $\Lambda$CDM scenario as the broadly accepted concordance model of modern cosmology. In spite of its simplicity and ability to be in tandem with a wide variety of up to date observational data \cite{Betoule:2014frx,Aubourg:2014yra,Baxter:2016ziy,Alam:2016hwk,Efstathiou:2017rgv,Pan-STARRS1:2017jku,Planck:2018vyg}, there are still theoretical and observational challenges that can not be explained in the context of the \lcdm paradigm. The theoretical challenges include the well known cosmological constant \cite{Weinberg:1988cp} and coincidence \cite{steinhardtbook,Velten:2014nra} problems. The observational challenges, which seem to have been amplified during the last decade, refer to the discrepancies between the reported values of some basic cosmological parameters using different cosmological probes/methods.

Probably the most important observational crack in the standard scenario is the \textit{$H_0$ tension}, which refers to the inconsistency of the Hubble parameter when comparing its value, obtained by the Cosmic Microwave Background (CMB) and Baryon Acoustic Oscillations (BAO) data, with the local Type Ia Supernovae (SnIa) measurements. The former use the inverse distance ladder method and lead to $H_0=67.4\pm 0.5\;km\,sec^{-1}$ Mpc$^{-1}$~\cite{Planck:2018vyg}, while the latter give $H_0=73.2\pm1.3\;km\,sec^{-1}$ Mpc$^{-1}$ by utilizing the distance ladder method~\cite{Riess:2020fzl}. This discrepancy is at $4\sigma$ level and reaches the $5\sigma$ threshold, when taking into account the latest measurement of the Pantheon+ dataset, suggesting $H_0=73.04\pm1.04\;km\,sec^{-1}$ Mpc$^{-1}$~\cite{Riess:2021jrx}. Another similar inconsistency, though at a somewhat lower level, is the \textit{growth tension}. The latter describes the inconsistency of the density rms matter fluctuations (in spheres of radius around $8h^{-1}\,$ Mpc) and/or of the matter density parameter ($\sigma_8$ and $\Omega_{0m}$ respectively), between the values reported by the Planck mission~\cite{Planck:2018vyg} and those reported by dynamical probes, such as Weak Lensing~\cite{Joudaki:2017zdt,DES:2017myr,Heymans:2020gsg} and the Redshift Space Distortion data~\cite{Macaulay:2013swa,Nesseris:2017vor,Kazantzidis:2018rnb,Perivolaropoulos:2019vkb,Skara:2019usd}. This mismatch is currently at a $2-3\sigma$ level.

These challenges led throughout the years to a wide variety of alternative theories that have the potential to address some of the issues discussed above (see Refs.~\cite{Ishak:2018his,Kazantzidis:2019dvk,DiValentino:2021izs,Perivolaropoulos:2021jda,CANTATA:2021ktz} for recent reviews on the subject and references therein). However, a common assumption in the majority of the proposed alternative scenarios is that any effects that may emerge from the observed large-scale peculiar motions are neglected. As a matter of fact, the authors in Ref. \cite{Davis:2010jq} have argued that neglecting coherent velocities in the SnIa sample can have a small but measurable impact on the estimation of cosmological parameters. Peculiar motions are induced by the gravitational attraction of the surrounding matter and lead to deviations from the smooth Hubble law. As a result, peculiar velocities are used as probes to test different cosmological models.  

Since the measurements of the individual peculiar velocities of galaxies still present large systematic uncertainties, the studies have been conducted on peculiar-velocity statistics, averaging over many galaxies. By far the most common is the \textit{bulk flow} (sometimes also called \textit{streaming motion}). This is a sensitive probe of the density fluctuations on large enough scales,  representing the coherent motion of a large region (usually a sphere centered on the observer) moving relative to the cosmological rest-frame (where the CMB temperature dipole vanishes). Over the last couple of decades, several studies have focused on the analysis of the peculiar-velocity field in search of an overall bulk flow. These studies use data from SnIa ~\cite{Riess:1995cg,Colin:2010ds,Dai:2011xm,Turnbull:2011ty,Feindt:2013pma,Lavaux:2012jb,Rathaus:2013ut,Ma:2013oja,Boruah:2019icj,2021arXiv211103055H}, employ galaxies as tracers~\cite{Feldman:2009es,Nusser:2011tu, Nusser:2011sd,Ma:2012tt,Hong:2014jla,Hoffman:2015waa, Magoulas,Scrimgeour:2015khj, Qin:2018sxw, Qin:2021jcs,Howlett:2022len}, or appeal to the kinematic Sunyaev-Zeldovich effect on the CMB, in order to measure bulk peculiar velocities of more than 1000 galaxy clusters~\cite{Kashlinsky:2008ut,Kashlinsky:2009dw, Mak:2011sw}. While most studies agree on the general direction of the bulk flow, there are strongly conflicting claims as to the amplitude and its scale. The majority of the surveys report bulk flows on scales $20-100h^{-1}$ Mpc and velocities around $200 - 300\;km\,sec^{-1}$~\cite{Ma:2013oja,Boruah:2019icj,Qin:2021jcs}. Also, although bulk peculiar flows should generally diminish on progressively larger volumes, there have been some controversial claims of very large bulk flows that are commonly referred to as \textit{dark flows}. These can exceed $400h^{-1}$ Mpc~\cite{Kashlinsky:2008ut} and are inconsistent with the \lcdm expectations. Alleviating this inconsistency may provide us with further insight into the nature of dark matter and dark energy.

Typically, most studies measure the amplitude and the direction of the bulk flow in the rest-frame of the CMB radiation, which also defines the coordinate system of the smooth Hubble flow. However, real observers do not simply follow the cosmological expansion, but drift relative to it, having their own peculiar motion.  For example, from the dipole anisotropy seen in the CMB temperature map, it is estimated that the Local Group of galaxies (including our Milky Way) has a peculiar velocity of about $600\;km\,sec^{-1}$ relative to the CMB rest-frame~\cite{Kogut:1993ag}. 

The present analysis adopts the tilted cosmological scenario introduced in Refs.~\cite{Tsagas:2009nh, Tsagas:2011wq}, which allows for two families of observers that move with respect to each other. This model predicts different values for the deceleration parameters measured in the two frames, due to relative-motion effects alone. In particular, observers residing inside locally contracting bulk flows may assign negative values to their local deceleration parameter, while the host universe is still globally decelerating. The apparent acceleration experienced by these observers is not real, but a local artefact of their peculiar motion. Nevertheless, the affected scales are large enough (typically between few hundred and several hundred Mpc) to create the false impression of a recent global event~\cite{Tsagas:2015mua,Tsagas:2021ldz}. Given that drift flows introduce apparent (Doppler-like) dipolar anisotropies, the aforementioned observers should also ``see''  an apparent dipole in the sky-distribution of the deceleration parameter~\cite{Tsagas:2011wq}. Moreover, the associated dipole axis should lie fairly close to that of CMB, assuming that both dipoles are (Doppler-like) relative-motion effects. Interestingly, over the last decade or so, a number of reports have claimed that such a dipolar anisotropy may actually exist in the local SnIa data~\cite{2010MNRAS.401.1409C,Antoniou:2010gw,Cai:2011xs,Mariano:2012wx,Mariano:2012ia,Colin:2019opb, Krishnan:2021jmh}. Similar dipolar signatures have also been reported using galaxy-cluster surveys. In the recent studies of \cite{Migkas:2021zdo,2021A&A...649A.151M}, for example, the authors reported a dipolar anisotropy in the distribution of the local Hubble constant ($H_0$), with a statistical significance of more than $5\sigma$, which could be attributed to a bulk flow of $900\; km \; sec^{-1}$ extending out to $\sim 500 \,$ Mpc. Similar variations on $H_{0}$ across the sky have been reported by \cite{Luongo:2021nqh} where QSOs and GRBs where used to trace the anisotropies. Additional hints for possible deviation from isotropy were recently claimed by using $1.36$ million quasars~\cite{Secrest:2020has}. The latter study revealed a large dipole well aligned with that of the CMB, showing a $4.9\sigma$ tension with respect to the one expected by standard $\Lambda$CDM. All these reports constitute a crucial test for the validity of the concordance $\Lambda$CDM paradigm.

In the current study we focus on the implications of bulk peculiar flows for the mean kinematics of the associated observers, by employing the latest publicly available SnIa (\ie the Pantheon compilation). More specifically, we provide the profile of the deceleration parameter in tilted almost-Friedmann universes, as measured by observers living inside (slightly) contracting bulk flows. This follows after adopting a functional form for the volume scalar, which monitors the local contraction of the observers' peculiar motion and in so doing determines the local deceleration parameter. Here, our results apply to a perturbed, tilted Einstein-de Sitter universe, which was selected for its naturalness and simplicity.\footnote{The relative-motion effects analysed here are not exclusive to the Einstein-de Sitter background, but apply to essentially all Friedmann-Robertson-Walker (FRW) backgrounds, irrespective of their spatial curvature and equation of state, including all those with matter that satisfies the strong-energy condition~\cite{Tsagas:2021tqa}. Also note that the effects are purely general relativistic in nature and cannot be naturally reproduced by a Newtonian study~\cite{Tsagas:2021dsl}.} In such a model, the accelerated expansion does not require an ad hoc cosmological constant, or exotic dark energy, but it is achieved naturally after allowing for peculiar-velocity perturbations and by introducing tilted observers. Clearly, allowing for bulk peculiar motions is far less arbitrary and has much sounder physical motivation than introducing a fine-tuned cosmological constant, or appealing to an elusive dark-energy component. In addition, since peculiar velocities are triggered by structure formation, the tilted scenario faces no coincidence problem either.

This work ameliorates previous treatments (see~\cite{Tsagas:2011wq, Tsagas:2015mua,Tsagas:2021ldz}), which only qualitatively constrained the parameters of the tilted model. The aforementioned earlier studies did not include a detailed data analysis, but used parameters such as the value of the deceleration parameter at $z=0$, the local expansion rate ``experienced'' by the real observers and the ``transition redshift'' from global deceleration to (apparent) local acceleration.

In the context of this analysis we raise and address the following two questions:
\begin{itemize}
    \item Can the deceleration parameter of a tilted cosmological model achieve a reasonable fit to the Pantheon data set, without the assumption of a cosmological constant or of dark energy?
    \item What are the favored values of the tilted parameters and how does the data-fit of the tilted scenario compares to that of the standard \lcdm paradigm, in the context of model selection criteria?
\end{itemize}
The structure of our paper is as follows: In the next section we present the basic theory describing the linear kinematics of relatively moving observers and provide the deceleration parameter measured by the tilted (\ie the real) observers. Section~\ref{sec:panthanalysis} contains the statistical analysis and the corresponding quality of fit, employing the Pantheon data set. Finally, in Section \ref{sec:concl}, we outline our results and discuss possible extensions and future work.

\section{Form of the Peculiar Deceleration Parameter}\label{sFPDP}
Following~\cite{Tsagas:2009nh,Tsagas:2011wq,Tsagas:2015mua, Tsagas:2021ldz} and adopting natural units (with $\hbar=c=8\,\pi\,G\equiv1$), we consider a tilted cosmological model equipped with two families of relatively moving observers. The first are the idealised (fictitious) observers, which follow the smooth Hubble expansion of a dust-dominated FRW universe with worldlines tangent to the timelike 4-velocity $u_{a}$ (normalised so that $u^{a}u_{a} = -1$).\footnote{Hereafter, we will treat the $u_{a}$-field as the reference frame of the universe, with respect to which large-scale peculiar velocities can be defined and measured.} The second family are the real (the so-called tilted) observers, residing in a typical galaxy like our Milky Way and drifting relative to the reference $u_a$-frame with finite peculiar velocity. The latter is non-relativistic, as indicated by the observations~\cite{Hong:2014jla, Boruah:2019icj,Qin:2021jcs}. Assuming that $\tilde{u}_a$ and $\Tilde{\upsilon}_{a}$ are respectively the 4-velocity and the peculiar velocity of the tilted observers (with $\tilde{u}_a\tilde{u}^a=-1$, $u_a\tilde{\upsilon}^a=0$ and $\tilde{\upsilon}^2 =\Tilde{\upsilon}_{a} \Tilde{\upsilon}^{a}\ll 1$), the aforementioned three velocity fields are related by the reduced Lorentz boost
\be
\Tilde{u}_{a} = u_{a} + \Tilde{\upsilon}_{a}\,.  \label{eq:veloc}
\ee
Note that the peculiar velocity vanishes in the FRW background by default, which makes $\Tilde{\upsilon}_{a}$ a gauge-invariant linear perturbation~\cite{Tsagas2007RelativisticCA, ellis_maartens_maccallum_2012}.\footnote{If $g_{ab}$ represents the metric of the whole spacetime, the symmetric spacelike tensors $h_{ab}=g_{ab}+u_{a}u_{b}$ and $\Tilde{h}_{ab}=g_{ab}+\Tilde{u}_{a}\Tilde{u}_{b}$ (with $h_{ab}u^{b} = 0 = \Tilde{h}_{ab}\Tilde{u}^b$ and $h_{a}\;^{a} = \Tilde{h}_{a}\;^{a} = 3$) project orthogonally to $u_{a} $ and to the  $\Tilde{u_{a}}$ fields respectively, while they also act as the metric tensors of the corresponding 3-spaces~\cite{Tsagas2007RelativisticCA, ellis_maartens_maccallum_2012}.}

In what follows, we will refer to the reference coordinate system of the ideal observers as the Hubble-frame, or the CMB-frame, while to that of the tilted observers as the bulk-flow frame or the tilted-frame. We will also adopt the relativistic $1+3$ covariant approach to cosmology (see \cite{Ehlers:1961xww,ellis1973cargese} and also \cite{Tsagas2007RelativisticCA,ellis_maartens_maccallum_2012} for more recent reviews), where spacetime splits into a temporal direction and into 3-dimensional space-like slices (along and orthogonal to the observers 4-velocity).

The mean kinematics of the relatively moving observers are determined by the volume scalars of their motion, monitoring the expansion/contraction of the corresponding velocity fields. Taking the spatial divergence of \eqref{eq:veloc} and keeping up to first-order terms leads to~\cite{Maartens:1998qw}
\be 
\Tilde{\Theta} = \Theta + \Tilde{\theta}\,, \label{eq:thetas}
\ee
where $\Theta={\rm D}^au_a$, $\tilde{\Theta}=\tilde{\rm D}^a\tilde{u}_a$ and $\tilde{\theta}=\tilde{\rm D}_a\tilde{\upsilon}_a$ are the associated expansion/contraction scalars.\footnote{Throughout this manuscript ``tildas'' will denote variables and operators evaluated in the tilted frame of the real observers. Thus, the differential operators ${\rm D}_a=h_a{}^b\nabla_b$ and $\tilde{\rm D}_a=\tilde{h}_a{}^b\nabla_b$, with $\nabla_a$ representing the spacetime covariant derivative operator, indicate spatial covariant derivatives in the Hubble and the tilted frames respectively.} Although the first two scalars are always positive due to the universal expansion, the third can be either positive or negative (\ie $\tilde{\theta}\gtrless0$, with $|\tilde{\theta}|/\Theta\ll1$ during the linear phase), since the bulk flow may locally expand or contract.\footnote{By construction, the expansion scalar is related to the average Hubble parameter ($H$) by means of the simple relation $\Theta=3H$, where $H=\dot{a}/a$ and $a=a(t)$ is the cosmological scale factor~\cite{Tsagas2007RelativisticCA,ellis_maartens_maccallum_2012}. Here, for convenience, we will also express the Hubble parameter in terms of redshift (\ie we will also adopt the notation $H=H(z)$). Finally, when referring to the current Hubble constant, we will use the notation $H_0$.} Following (\ref{eq:thetas}), observers inside contracting bulk flows (where $\Tilde{\theta}<0$) will measure a smaller expansion rate than their Hubble-flow counterparts (\ie $\Tilde{\Theta}\lessapprox \Theta$). Inside a locally expanding bulk flow (where $\Tilde{\theta}>0$), on the other hand, the tilted observers will ``experience" faster expansion (with $\Tilde{\Theta}\gtrapprox \Theta$). Nevertheless, in either case, the relative-motion effect should be too small to leave a measurable imprint, since $|\tilde{\theta}|/\Theta\ll1$ at the linear level.

According to (\ref{eq:thetas}), the expansion rates measured in the Hubble and the tilted frames differ simply because of relative-motion effects. For the same reason, the two observer groups experience different acceleration/deceleration rates as well. Indeed, differentiating Eq.~\eqref{eq:veloc} with respect to time and keeping up to first-order terms, leads to the linear relation
\be 
\Tilde{\Theta}^{\prime} = \dot{\Theta} + \Tilde{\theta^{\prime}}\,, \label{eq:exprates}
\ee
where $\dot{\Theta}$ and $\Tilde{\Theta}^{\prime}$ monitor the time evolution of $\Theta$ and $\Tilde{\Theta}$ in their respective rest frames~\cite{Tsagas:2009nh}.\footnote{Hereafter, overdots will indicate time differentiation in the Hubble/CMB-frame, while primes will denote the same in the tilted coordinate system of the real observers. In other words, $\dot{\Theta}= u^{a}\nabla_{a}\Theta$, $\Tilde{\Theta}^{\prime}= \tilde{u}^{a}\nabla_{a}\Tilde{\Theta}$ and $\Tilde{\theta}^{\prime}= \tilde{u}^{a}\nabla_{a}\Tilde{\theta}$~\cite{Tsagas:2009nh,Tsagas:2011wq}.} At this point we should note that, although $|\tilde{\theta}|/\Theta\ll1$ all along the linear phase, this is not necessarily true for the ratio of their time derivatives.

The linear differences between the volume scalars and between their time derivatives, seen in (\ref{eq:thetas}) and (\ref{eq:exprates}), imply that the deceleration parameters measured in the CMB and the tilted frames differ as well. Indeed, expressed in terms of their volume scalars and their derivatives, the deceleration parameters measured in the coordinate system of the smooth Hubble flow and in rest-frame of the bulk peculiar motion read
\begin{equation}
q= -\left(1 +\frac{3\dot{\Theta}}{\Theta^{2}}\right)  \label{q}
\end{equation}
and
\begin{equation}
\Tilde{q} = -\left(1 + \frac{3\Tilde{\Theta}^{\prime}}{\Tilde{\Theta}^{2}} \right)\,, \label{eq:decparams}
\end{equation}
respectively. Combining these definitions with Eqs.~(\ref{eq:thetas}) and (\ref{eq:exprates}), leads to the following linear relation between the two deceleration parameters~\cite{Tsagas:2009nh,Tsagas:2011wq, Tsagas:2015mua,Tsagas:2021ldz}
\be 
\Tilde{q} = q - \frac{\tilde{\theta}'}{3H^{2}}= q + \frac{1}{3}\left(1 + \frac{\Omega}{2}\right)\frac{\Tilde{\theta}^{'}}{\dot{H}}\,,
\label{eq:q_omega}
\ee
guaranteeing that $\tilde{q}\neq q$ due to the presence of peculiar motions. Note that, in deriving the above, we have also used the background expressions $\Theta=3H$ and $\dot{H}=-H^{2}\left(1+\Omega/2 \right)$, with $\Omega= \kappa\rho/3H^{2}$ representing the density parameter of an FRW universe. 

The second term on the right-hand side of \eqref{eq:q_omega} constitutes a correction term introduced by relative-motion effects, since it vanishes in the absence of them. In order to analyse Eq.~\eqref{eq:q_omega} further, we need an expression for the time evolution of $\tilde{\theta}$ written in the bulk-flow frame. Put another way, we need the Raychaudhuri equation of the bulk peculiar motion. In a perturbed, tilted FRW universe with dust, the latter linearises to~\cite{Tsagas:2013ila} 
\be 
\tilde{\theta}^{'} = - H \tilde{\theta} + \tilde{\rm D}^{a}\tilde{\upsilon}_{a}^{'}\,.  \label{eq:raychaud}
\ee
The source term in the above, namely the spatial divergence of the time derivative of the peculiar velocity ($\tilde{\rm D}^a\tilde{\upsilon}^{'}_a$), is not yet directly observable and  requires further theoretical analysis. We will therefore turn to linear relativistic cosmological perturbation theory.

Relative motions also affect the nature of the cosmic medium ``seen'' by the associated observers. Following~\cite{Maartens:1998qw}, the linear relations between the matter components measured in the two frames are $\tilde{\rho}=\rho$, $\tilde{p}=p$, $\tilde{q}_a=q-(\rho+p)\tilde{\upsilon}_a$ and $\tilde{\pi}_{ab}=\pi_{ab}$. Here, $\rho$ is the energy density, $p$ is the (isotropic) pressure, $q_a$ is the energy flux and $\pi_{ab}$ is the viscosity of the matter. Accordingly, when the Hubble-flow observers see the cosmic medium as a pressureless perfect fluid (with $p=0=q_a=\pi_{ab}$), the tilted see it as imperfect, with an effective energy flux $\tilde{q}_a=-\rho\tilde{\upsilon}_a$ due to relative-motion effects alone. 

In contrast to Newtonian gravity, the energy flux also contributes to the relativistic gravitational field, via the local energy-momentum tensor. In a sense, one may say that, in relativity, bulk flows gravitate~\cite{2020EPJC...80..757T,Filippou:2020gnr}. When dealing with bulk peculiar flows, where there is a nonzero $\tilde{q}_a$ due to the observers relative motion, this additional flux-contribution to the local gravitational field acquires particular significance.

The extra input of the bulk-flow flux to the Einstein field equations feeds into the relativistic conservation laws and eventually emerges in the equations monitoring the linear evolution of the tilted cosmological model. More specifically, in a tilted almost-FRW universe with dust, expression (\ref{eq:raychaud}) recasts into~\cite{Tsagas:2021ldz,Tsagas:2021dsl}
\be 
\tilde{\theta}^{'} = - 2H \tilde{\theta} + \frac{1}{3H} \tilde{\rm D}^{2}\tilde{\theta} - \frac{1}{a^{2}} \left(\frac{\tilde{\Delta}^{'}}{3H} + \frac{\tilde{\mathcal{Z}}}{3H} \right)\,. \label{eq:raychaud2}
\ee
Note that $\Delta$ describes inhomogeneities in the spatial distribution of the matter density and $\mathcal{Z}$ represents spatial variations in the universal expansion.\footnote{By construction $\tilde{\Delta}=a\tilde{\rm D}^{a}\tilde{\Delta}_{a}$, with $\tilde{\Delta}_a=(a/\rho)\tilde{\rm D}_a\rho$. The former closely corresponds to the familiar density contrast $\delta=\delta\rho/\rho$~\cite{Tsagas2007RelativisticCA,ellis_maartens_maccallum_2012}. Also by construction, $\tilde{\mathcal{Z}}=a\tilde{\rm D}^{a}\tilde{\mathcal{Z}}_{a}$, with $\tilde{\mathcal{Z}}= a\tilde{\rm D}_a\tilde{\Theta}$.} For more details and further discussion, the reader is referred to~\cite{Tsagas:2021ldz, Tsagas:2021dsl}.

Before substituting (\ref{eq:raychaud2}) into Eq.~(\ref{eq:q_omega}), it is worth noting the spatial Laplacian term on the right-hand side of (\ref{eq:raychaud2}). The latter introduces a scale-dependence, which becomes explicit after a simple harmonic decomposition. In so doing, one arrives at the following expression 
\be
\tilde{q}_{(n)} = \; q + \frac{1}{9} \left (\frac{\lambda_{H}}{\lambda_{(n)}}  \right)^{2} \frac{\tilde{\theta}_{(n)}}{H}+\frac{1}{9} \left ( \frac{\lambda_{H}}{\lambda_{K}}\right)^{2}\left(  \frac{\tilde{\Delta}^{'}_{(n)}}{H}  + \frac{\tilde{\mathcal{Z}}_{(n)}}{H}  \right)\,, \label{eq:q_pert}
\ee
for the $n$-th harmonic~\cite{Tsagas:2021ldz, Tsagas:2021dsl}. Here, $\lambda_{(n)}$ is the physical scale of the peculiar-velocity perturbation (practically the size of the bulk flow), $\lambda_{H}=1/H$ is the Hubble radius (quantifying the size of the observable universe, with $\lambda_{H}\thickapprox 4300$~Mpc today) and $\lambda_K = a/|K|$ is the curvature scale of the FRW background (with $K=\pm1$). Since $(\lambda_{H}/\lambda_K)^2=|1-\Omega|$ and given that $|1-\Omega|\le10^{-2}$, due to the near spatial flatness of our cosmos~\cite{Planck:2018vyg}, the last term of \eqref{eq:q_pert} is negligible. The latter then simplifies to 
\be 
\Tilde{q} = q + \frac{1}{9} \left (\frac{\lambda_{H}}{\lambda}  \right)^{2} \frac{\tilde{\theta}}{H}\,, \label{eq:q_lambda1}
\ee
after dropping the mode index ($n$) for economy. The above provides a very simple relation between $\tilde{q}$ and $q$. The former is the local deceleration parameter measured in the rest-frame of a bulk flow with size $\lambda$. The latter is measured in the coordinate system of the smooth Hubble expansion and coincides with the deceleration parameter of the universe itself. Note that, even though $|\tilde{\theta}|/H\ll1$ throughout the linear regime, the effect of the (purely relativistic) correction term on the right-hand side of Eq.~\eqref{eq:q_lambda1} can be strong, depending on the scale of the bulk flow in question. In fact, the relative-motion effect on $\tilde{q}$ increases as we move down to smaller scales, where $\lambda_{H}/\lambda\gg1$. Furthermore, the overall effect also depends on the sign of $\tilde{\theta}$. This means that in slightly expanding bulk flows (where $0<\tilde{\theta}/H\ll1$) the deceleration parameter measured in the tilted frame is greater than its Hubble-frame counterpart. In contrast, the deceleration parameter measured by observers inside slightly contracting bulk flows (\ie those with $-1\ll\tilde{\theta}/H\ll0$) is smaller than the deceleration parameter of the universe.

Following \eqref{eq:q_lambda1}, the local value of $\tilde{q}$ is more sensitive to the scale-ratio $\lambda_{H}/\lambda$. This means that on super-Hubble lengths (with $\lambda_{H}/\lambda\ll1$) the impact of the observer's relative motion is negligible and the local deceleration parameter  approaches its global value (\ie $\tilde{q}\rightarrow q$). This also agrees with our expectation that the peculiar velocities and their effects fade away as we move out to progressively longer wavelengths. 

The situation changes drastically on sufficiently small scales, where $\lambda_{H}/\lambda\gg1$. There, even relatively slow bulk flows (with $|\tilde{\theta}|/H\ll1$) can have a significant impact on the local deceleration parameter. In fact, the relative-motion effects dominate the right-hand side of \eqref{eq:q_lambda1} when the correction term equals (in absolute value) the deceleration parameter ($q$) measured in the Hubble frame. This occurs at a characteristic length given by~\cite{Tsagas:2021ldz}
\be 
\lambda_{T} = \sqrt{\frac{1}{9q}\frac{|\tilde{\theta}|}{H}} \lambda_{H}\,. \label{eq:Jeansl}
\ee
On scales smaller than $\lambda_T$, peculiar-velocity perturbations dominate over the background Hubble expansion, dictating the linear bulk-flow kinematics and determining the local value of the deceleration parameter.\footnote{It is worth pointing out the close analogy between $\lambda_T$ and the familiar ``Jeans length'' ($\lambda_J$). Recall that the latter determines the threshold inside which pressure-gradient perturbations dominate over the background gravitational pull and thus dictate the linear evolution of density perturbations. In this respect, $\lambda_T$ can be seen as the peculiar-motion analogue of the Jeans length, or as the ``peculiar Jeans length''~\cite{Tsagas:2021ldz}.} The latter can even turn negative when the bulk flow is contracting. In that case, $\lambda_T$ also marks the ``transition scale'' where the sign of $\tilde{q}$ changes from positive to negative~\cite{Tsagas:2021tqa,Tsagas:2021ldz}. Using data from recent bulk-flow surveys, the typical values of $\lambda_T$ were found to vary between few hundred and several hundred Mpc (see~\cite{Tsagas:2021ldz} for a representative Table).  

Focusing on contracting bulk flows (with $\tilde{\theta}<0$), Eqs.~\eqref{eq:q_lambda1} and \eqref{eq:Jeansl} combine to give \cite{Tsagas:2021dsl}
\be 
\Tilde{q} =  q\left[1-\left(\frac{\lambda_{T}}{\lambda} \right)^2\right]\,,  \label{eq:q_lambda2}
\ee
guaranteeing that $\tilde{q}<0$ on scales smaller than the transition length (with $\lambda<\lambda_T$). Accordingly, observers inside (slightly) contracting peculiar motions will assign negative values to their local deceleration parameter within a region determined by the associated transition length. On scales larger than $\lambda_T$, on the other hand, the same observers will assign positive values to $\tilde{q}$. Clearly, on sufficiently large wavelengths, the correction term on the rght-hand side of (\ref{eq:q_lambda2}) becomes negligible and the value of the local deceleration parameter approaches that of its Hubble-flow counterpart (\ie $\tilde{q}\rightarrow q$). 

It is therefore theoretically possible to achieve accelerated expansion in tilted almost-FRW universes without appealing to dark energy, or to a cosmological constant. There is also no need to modify general relativity, to abandon the Friedmann models, or to introduce any new physics. Everything takes place within standard cosmology and conventional physics. One only needs to allow for large-scale peculiar motions, like those reported by an ever increasing number of surveys.

Clearly, the accelerated expansion ``experienced'' by the bulk-flow observers is not real, but a local artefact of their peculiar motion relative to the smooth Hubble expansion. Globally, the host universe is still decelerating with $q>0$. Nevertheless, the affected scales are large enough ($\lambda_T$ typically ranges from few hundred to several hundred Mpc) to create the false impression of a recent global event. Put another way, the unsuspecting observers have misinterpreted the
local contraction of the bulk flow they happen to reside in, as global acceleration of the surrounding universe.

Looking back at Eqs.~(\ref{eq:q_lambda1}) and \eqref{eq:Jeansl}, one can easily see the key role played by the local volume scalar ($\tilde{\theta}$) in determining the values of both $\tilde{q}$ and $\lambda_T$. For simplicity, we have so far implicitly assumed that $\tilde{\theta}$ remains constant within the whole of the bulk-flow domain. In practice, however, one expects the local volume scalar to vary with scale. We therefore need to consider physically motivated profiles for $\tilde{\theta}$ and then test their fit to the observations.

Given that peculiar velocities fade away with increasing scale, we expect the value of $\tilde{\theta}$ to decrease as we move out to progressively larger wavelength (\ie $\tilde{\theta}\rightarrow0$ as $\lambda\rightarrow\infty$). Also, when dealing with contracting bulk motions, we expect faster contraction rates near the outskirts of the flow and slower towards its center. Physically, this is a well motivated pattern, since it agrees with the typical kinematic behaviour of any contracting self-gravitating system. The above described qualitative profile of $\tilde{\theta}$ is well parametrised by the functional form
\be 
\Tilde{\theta}= \tilde{\theta}(\lambda)= \frac{m\,\lambda^2}{p+r\,\lambda^3}\,,   \label{eq:volscal}
\ee
where $m$, $p$ and $r$ are parameters decided by the data. Finally, given that the background Hubble parameter remains essentially constant (\ie $H\simeq H_0$) within the scales/redshifts of interest, Eqs.~\eqref{eq:volscal} and \eqref{eq:q_lambda2} lead to the following functional form
\be
\tilde{q}= \Tilde{q}(\lambda)= \frac{1}{2}\left(1-\frac{m}{p +r\,\lambda^3}\right)\,,  \label{eq:q_lambda3}
\ee
for the deceleration parameter measured by the real observers. Note that we have set $q=1/2$ in the above, which implies that expression (\ref{eq:q_lambda3}) holds on an Einstein-de Sitter background. Having said that, we remind the reader that the above result applies to essentially all FRW backgrounds, irrespective of their spatial curvature and equation of state (see~\cite{Tsagas:2021dsl} and also footnote~1 here).

\begin{centering}
\begin{table*}
\caption{Summary of Sn1a subsamples of the Pantheon dataset. The column $N_{SnIa}$ includes the total number of SnIa of every sample and the two last columns indicate the median value of the CMB redshift and its coverage for each sample respectively. The corresponding $N_{SnIa}$ number were identified using the idsurvey column from the \texttt{Ancillary\_C11.FITRES} file of the Pantheon data in the corresponding github repository.}
\label{table:compilation}
\resizebox{0.6\textwidth}{!}{
\begin{tabular}{|c|c|c|c| }
\hline
Sample & $N_{SnIa}$ & Redshift Median & Redshift Range \\
\hline 
\rule{0pt}{3ex}
CfA 1-4 & 147 & 0.03 & 0.01 - 0.07\\
CSP  & 25 & 0.02 & 0.01 - 0.06 \\
SDSS & 335 & 0.20 & 0.03 - 0.40  \\
SNLS &236  & 0.64 & 0.12 - 1.06 \\
PS1 & 279& 0.29 & 0.02 - 0.63 \\
high-$z$ & 26& 1.26 & 0.73 - 2.26 \\
\hline
all &1048 &0.25 & 0.01 - 2.26 \\
\hline
\end{tabular}
}
\end{table*}
\end{centering}

\begin{figure*}
\centering
\includegraphics[width =1.\textwidth]{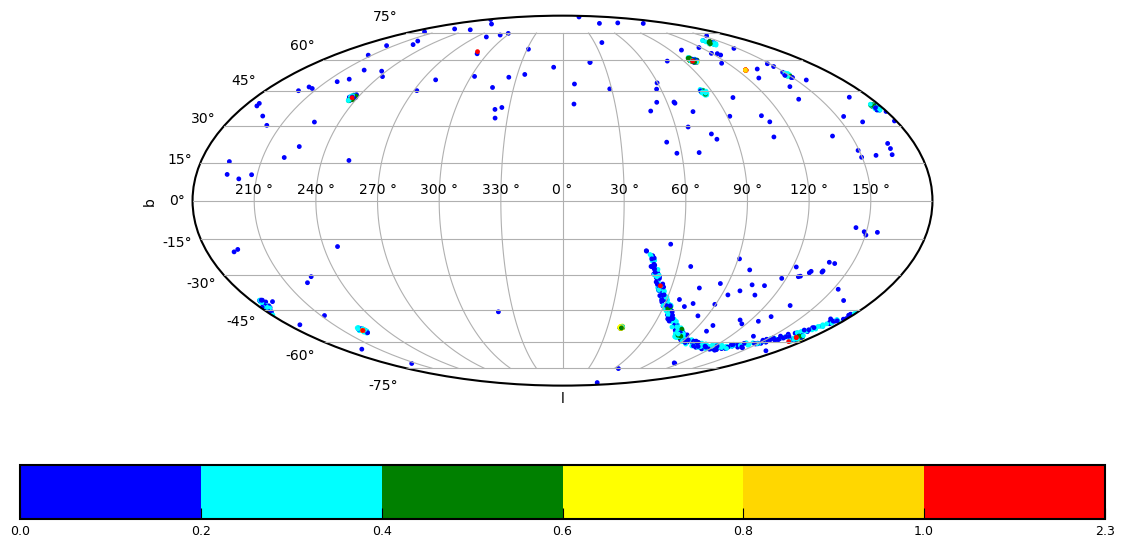}
\caption{The sky distribution of the Pantheon SnIa sample in galactic coordinates classified by redshifts. The pseudo-colours indicate the supernovae redshifts with respect to the CMB frame.}
\label{fig:Mollweide}
\end{figure*} 

\section{Pantheon S${\rm n}$I${\rm a}$ Analysis}
\label{sec:panthanalysis}
In order to constrain the parameters $m$, $p$ and $r$ that appear in the tilted deceleration parametrization \eqref{eq:q_lambda3}, we use the latest publicly available SnIa sample, namely the Pantheon sample consisting of 1048 spectroscopically confirmed SnIa originally compiled by ~\cite{2018ApJ...859..101S}.\footnote{Actually, the latest SnIa sample corresponds to the Pantheon+ compilation \cite{Riess:2021jrx}. However, during the writing of the present work the data are not publicly available.}

\subsection{The Pantheon Supernovae Sample}
\label{Pantheon}
\begin{figure*}
\centering
\includegraphics[width =0.75\textwidth]{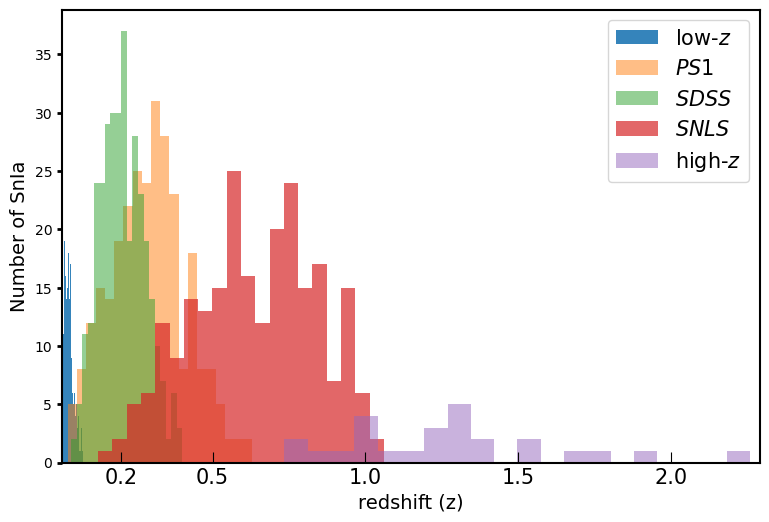}
\caption{The redshift distribution of the Pantheon SnIa subsamples, indicated in the legend, in the CMB frame. The blue colour indicates the number of SnIa in the CfA\, 1-4 and CSP samples, while with purple the SnIa from the  CANDELS/CLASH, GOODS and SCP subsamples are shown.}
\label{fig:z_distro}
\end{figure*}

The Pantheon sample corresponds to a compilation of different supernovae surveys detecting SnIa in both high and low redshifts regions. In particular, it incorporates the CfA1-CfA4 ~\cite{1999AJ....117..707R, 2006AJ....131..527J, 2009ApJ...700..331H, 2009ApJ...700.1097H, 2012ApJS..200...12H} surveys, the Pan-STARRS1 (PS1) Medium Deep Survey ~\cite{2018ApJ...859..101S}, the Sloan Digital Sky Survey (SDSS) ~\cite{2018PASP..130f4002S}, the SuperNovae Legacy Survey (SNLS) ~\cite{2010A&A...523A...7G}, ESSENCE ~\cite{2016ApJS..224....3N}, the Carnegie Supernova Project (CSP) ~\cite{2010AJ....139..519C} as well as various  Hubble Space Telescope (HST) samples, namely the CANDELS/CLASH ~\cite{2014ApJ...783...28G, 2014AJ....148...13R, 2018ApJ...853..126R}, GOODS ~\cite{2004ApJ...607..665R, 2007ApJ...659...98R} and SnIa form the Supernova Cosmology project (SCP) ~\cite{2012ApJ...746...85S} subsamples. A summary of these surveys is illustrated in the following Table \ref{table:compilation}, where we also present the redshift range that each survey covers as well as the total number of SnIa $(N_{SnIa})$. As a result the entire sample covers the redshift range $0.01 \leq z \leq 2.26$. The sky distribution (in galactic coordinates) of the SnIa is demonstrated in Fig. \ref{fig:Mollweide} where we have colour-coded each SnIa with respect to the redshift, while the redshift distribution of the entire sample is shown in Fig. \ref{fig:z_distro}. Notice that the sky distribution of the Pantheon sample is quite inhomogeneous, since the majority pf the SnIa lies in the ``southeast" quadrisphere.

The Pantheon sample contains significant improvements on the $PS1$ photometry to reduce the photometric calibration uncertainties to the milli-magnitude (mmag) level. The light curves of SnIa are parametrized by their brightness, observed colour and their decline rate. In ~\cite{2018ApJ...859..101S}, a recent version of light curve fitter is used, namely the Spectral Adaptive Lightcurve Template - 2 (SALT2, see Ref. ~\cite{2014A&A...568A..22B} for more details) in order to optimize the use of SnIa as standard candles to determine accurate distances and to reduce systematic uncertainties due to \textit{K}-corrections. Given the Tripp estimator ~\cite{1998A&A...331..815T} together with the light curve fitter, the standardized observational SnIa distance modulus is given by
\begin{equation}
    \mu_{obs} = m_b - M + \alpha \, x_{1} - \beta \, c + \Delta_{M} +\Delta_{B},
\end{equation}
where $m_b$ corresponds to the observed peak magnitude at time of B-band maximum, $M$ is the absolute B-band magnitude of a fiducial SnIa, $x_1$ is a light curve shape parameter, $c$ describes the SnIa colour at maximum brightness, $\alpha$ is a coefficient of the relation between luminosity and stretch and $\beta$ is a coefficient of the relation between luminosity and colour. Moreover, $\Delta_{M}$ and $\Delta_{B}$ are distance corrections based on the mass of the host galaxy and on predicted biases from simulations, respectively. As we can see, there are different sources of systematic uncertainties that can yield to inaccurate distance estimations. A distance-dependent bias that mostly affects the SnIa at cosmological redshift is associated with the Malmquist bias, where at redshifts near the survey magnitude threshold, brighter SnIa are most likely to be observed, biasing the effective luminosity towards higher values. In ~\cite{2018ApJ...859..101S}, the authors used the BEAMS with Bias Corrections (BBC) method ~\cite{2017ApJ...836...56K} to account for errors due to intrinsic scatter and selection effects (such as the the Malmquist bias), based on accurate SnIa simulations.

After all these corrections are applied, one can infer the corrected apparent magnitudes $m_{obs}$ and perform a cosmological fit based on a specific theoretical model. This approach is implemented in the next subsection.

\subsection{Pantheon Dataset Fit}
The SnIa are widely used as standard candles to probe the expansion rate, by utilizing the theoretically predicted apparent magnitude $m_b(z)$. The latter reads
\be 
m_b(z)=M+5 \, log_{10} \left[\frac{d_L(z)}{1Mpc} \right] +25\,, \label{eq:mbthdL}
\ee
where $M$ is the well known corrected intrinsic (absolute) magnitude with respect to the colour and stretch. Also, $d_L(z)$ corresponds to the luminosity distance, which in the context of a flat universe is calculated by
\be 
d_L(z)=c \, (1+z) \, \int_0^z \frac{dz'}{H(z')}\,,
\label{eq:dl}
\ee
where $z$ denotes the SnIa redshift in the CMB rest frame and $c$ is the speed of light. Typically, instead of the luminosity distance ($d_L$), the Hubble free luminosity distance ($D_L(z)\equiv H_0 \, d_L(z)/c$) is used for the theoretically predicted apparent magnitude, recasting Eq.~\eqref{eq:mbthdL} as
\be
m_{b}(z)= M +5 \, log_{10}\left[D_L(z)\right]+5\, log_{10}\left(\frac{c/H_0}{Mpc}\right)+25\,. \label{eq:mbthDL} 
\ee

From Eq. \eqref{eq:mbthDL} we clearly see a degeneracy between the parameters $M$ an $H_0$, which in the context of a \lcdm background $H(z)$ is considered constant. As a result, the two parameters are combined for the definition of the parameter $\cal{M}$, which is determined as  

\begin{figure*}
\begin{center}$
\begin{array}{cc}
\includegraphics[width= 0.5 \textwidth]{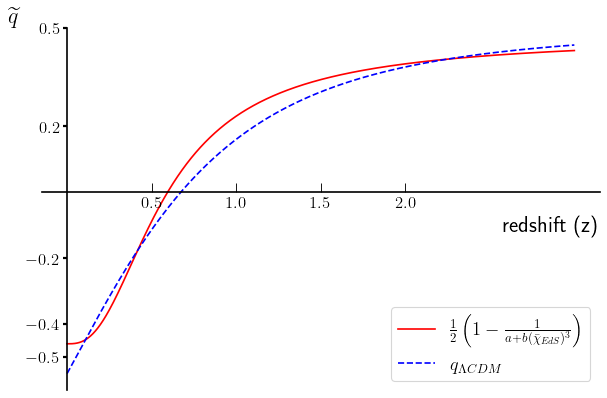}&
\includegraphics[width= 0.5 \textwidth]{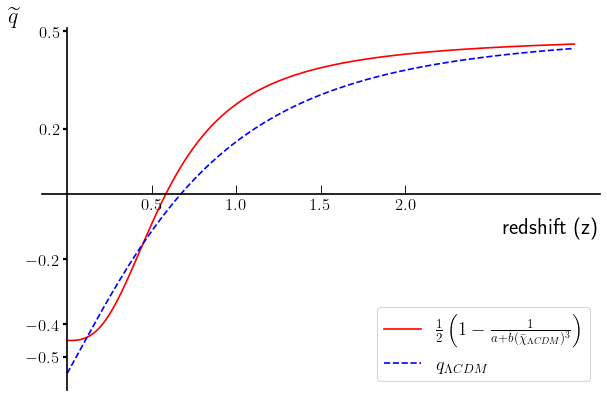}
\end{array}$
\end{center}
\caption{The evolution of $\Tilde{q}(\lambda(z)) = \frac{1}{2}\left(1-\frac{1}{\alpha+b \,\bar{\chi}_{EdS}^3}\right)$ (at the left panel) and $\Tilde{q}(\lambda(z)) = \frac{1}{2}\left(1-\frac{1}{\alpha+b \, \bar{\chi}_{\Lambda CDM}^3}\right)$ (at the right panel) by fitting it to the full Pantheon dataset. The red solid line is drawn by using the best fit parameters for each model (see Table \ref{table:bfres}). For comparison, the dashed blue line corresponds to the best fit \lcdm which has a deceleration parameter of the form $q_{\Lambda CDM}(z)=\left[\Omega_{0m}(1+z^3)-2(1-\Omega_{0m})\right]/\left[2(\Omega_{0m} (1+z)^3+1-\Omega_{0m})\right]$.}
\label{fig:q_distros}
\end{figure*}

\be
{\cal M} \equiv M +5 \, log_{10}\left[\frac{c/H_0}{1Mpc}\right]+25=M -5 \, log_{10}(h)+42.38\,,
\label{eq:Mcaldef} 
\ee
where $H_0=100h \;km \; sec^{-1}\,$ Mpc$^{-1}$. Even though this degenerate combination is marginalized in most cases \cite{SNLS:2011lii,Pan-STARRS1:2017jku}, recent studies \cite{Zhao:2019azy,Kazantzidis:2020tko,Sapone:2020wwz,Kazantzidis:2020xta,Dainotti:2021pqg,Dainotti:2022bzg} argue that this process may lead to physical information loss, since a physical model with an abrupt transition on the absolute magnitude $M$ at a low redshift $z_t$ has the potential to alleviate simultaneously the $H_0$ and growth tensions \cite{Alestas:2020zol,Camarena:2021jlr,Marra:2021fvf,Alestas:2021nmi,Perivolaropoulos:2021bds,Alestas:2021luu,Alestas:2022xxm,Perivolaropoulos:2022vql}. Hence, we decide to use this degenerate parameter in the minimization process.

Armed with Eqs. \eqref{eq:mbthdL}-\eqref{eq:Mcaldef}, we can apply the maximum likelihood method \cite{Arjona:2018jhh}, and construct the relevant $\chi^2$ function as follows
\be
\chi^2=V^i_{Panth.} \, C_{ij}^{-1} \, V^j_{Panth.}\,, \label{eq:chi2}
\ee
where $V^i_{Panth.}\equiv m_{obs}(z_i)-m_{b}(z)$ corresponds to the difference between the observed SnIa apparent magnitudes at redshift $z_{i}$ with the theoretically predicted ones calculated from Eq. \eqref{eq:mbthDL} and $C_{ij}^{-1}$ is the inverse of the total covariance matrix. The total covariance matrix is constructed by taking the sum of a diagonal matrix $D_{\textrm{stat}}$ that includes the statistical uncertainties  of the apparent magnitudes $m_{obs}(z_i)$ and a non diagonal matrix that is constructed using the systematic uncertainties due to the bias correction method (see Ref. \cite{Pan-STARRS1:2017jku} for more details). The diagonal matrix includes the total distance errors associated with every SnIa and takes the form
\[
D_{\textrm{stat}} = 
  \begin{pmatrix}
    \sigma^2_{m_{obs,1}} & 0 & \dots & 0 \\
    0 & \sigma^2_{m_{obs,2}} & \dots & 0 \\
    \vdots & \vdots & \ddots & \vdots \\
    0 & 0 & \dots & \sigma^2_{m_{obs,N}}
  \end{pmatrix}
\]
where 
$\sigma^2_{m_{obs}} = \sigma^2_{M} + \sigma^2_{Mass} + \sigma^2_{\mu - z} + \sigma^2_{lens} + \sigma^2_{int} + \sigma^2_{Bias}$. 
The individual uncertainty contributions to the matrix are the photometric error, the mass step correction, the peculiar velocity and redshift uncertainty in quadrature, the stochastic gravitational redshift, the intrinsic scatter and the distance bias correction respectively. From the definition of the luminosity distance, it is evident that in order to apply the maximum likelihood method we need a specific form for $H(z)$. From the deceleration parameter, we can construct the corresponding evolution of $H(z)$ via \cite{Gong:2006gs}
\begin{align} 
H(z)&=H_0 \, \exp \left[\int^z_0 \left[\frac{1+q(u)}{1+u} \right]du\right]\,. \label{eq:hubratqz}
\end{align}

\begin{centering}
\begin{table*}
\caption{Table of the best fit parameters for the two tilted cosmological models T-$\Lambda$ and T$-EdS$ as well as the standard $\Lambda$CDM scenario. Notice that with the acronyms T-$\Lambda$ and T$-EdS$ we refer to a tilted cosmological model with a \lcdm or an EdS line-of-sight comoving distance respectively. The tilted cosmological models have been fitted using Eqs. \eqref{eq:mbthdL}-\eqref{eq:hubratqz}. The three models in comparison giving similar $\chi^2_{min}$ values, using the Pantheon data.}
\label{table:bfres}
\resizebox{0.9 \textwidth}{!}{
\begin{tabular}{|c|c|c|c|c|c|c|c|c|}
\hline
Model  & $\cal{M}$ & $\alpha$ & $b$  & $\Omega_{0m}$ & $\chi_\textrm{min}^2$ & $\chi_\textrm{red}^2$\\
\hline\hline
\rule{0pt}{2.7ex} 
\textbf{$\mathbf{\Lambda}$CDM} & $\mathbf{23.809 \pm 0.011}$  & $\mathbf{-}$ &$\mathbf{-}$ & $\mathbf{0.299 \pm 0.022}$ & $\mathbf{1026.67}$ & $\mathbf{0.981}$\\
\rule{0pt}{2.7ex} 
T-$\Lambda$  & $23.815^{+0.014}_{-0.012}$ & $0.517^{+0.039}_{-0.038} $ & $3.9^{+3.6}_{-2.4}$ & $0.3$ & $1026.69$ & $0.982$\\
\rule{0pt}{2.7ex} 
T-$\Lambda$ ($\alpha$ fixed)  & $23.808 \pm 0.007$ & $0.5$ & $5.20 ^{+ 2.6}_{-1.9}$ & $0.3$ & $1027.21$ & $0.982$\\
\rule{0pt}{2.7ex} 
\textbf{T-$\mathbf{EdS}$}  & $\mathbf{23.813^{+0.015}_{-0.014}}$ & $\mathbf{0.512\pm 0.041}$  & $\mathbf{6.7^{+5.6}_{-3.8}}$ & $\mathbf{1.0}$ & $\mathbf{1026.76}$ & $\mathbf{0.982}$\\
\rule{0pt}{2.7ex} 
\textbf{T-$\mathbf{EdS}$ ($\alpha$ fixed)}  & $\mathbf{23.809\pm 0.007}$ & $\mathbf{0.5}$  & $\mathbf{8.56^{+3.8}_{-2.9}}$ & $\mathbf{1.0}$ & $\mathbf{1027.05}$ & $\mathbf{0.982}$\\
\hline
\end{tabular}}
\end{table*}
\end{centering}

\noindent Hence, for a specific form of the deceleration parameter, we can easily construct the theoretically predicted apparent magnitude $m_b(z)$ and as a result the corresponding $\chi^2$ function, solving Eq. \eqref{eq:hubratqz} for a specific $q(z)$ parametrization and substituting the derived formula of $H(z)$ to the luminosity distance $d_L(z)$ of Eq. \eqref{eq:dl}.

However, our parametrization \eqref{eq:q_lambda3} depends on the physical size of the bulk flow $\lambda$ and can not be used on the current form. So, in order to transform the deceleration parameter $\Tilde{q}$ to a redshift dependent function, we set $\lambda \equiv \bar{\chi}(z)$, where $\bar{\chi}(z)$ corresponds to the line-of-sight comoving distance. In the present analysis we choose two different cosmologies for the line-of-sight comoving distance. The standard \lcdm cosmology, where the line-of-sight comoving distance is given as 
\begin{flalign}
&\bar{\chi}_{\Lambda CDM}(z)= \int_0^z \frac{c \, dz'}{H_0 \, \sqrt{\Omega_{0m}(1+z')^3  +(1-\Omega_{0m}})}\,,&
\label{eq:comdistlcdm}
\end{flalign}
as well as the Einstein-de Sitter (EdS) form for the line-of-sight comoving distance ($\Lambda=0$) which is defined as
\be 
\bar{\chi}_{EdS}(z) = \frac{2c}{H_{0}}\left(1-\frac{1}{\sqrt{1+z}}\right)\,. \label{eq:comdisteds}
\ee

Moreover, we can further reduce the the number of parameters in Eq. \eqref{eq:q_lambda3} rearranging the constants $m,p$ and $r$ as follows
\be
\Tilde{q}(\lambda(z))=\frac{1}{2} \left[1-\frac{1}{\alpha+b\, d_r^{3}(z)} \right]\,, \label{eq:qzfull}
\ee
where $\alpha,b$ are dimensionless parameters of the tilted cosmological models to be fixed by the data and $d_r(z) \equiv H_0 \,\bar{\chi}(z)/c$. Recall that we have set  $\lambda \equiv \bar{\chi}(z)$ and depending on the bulk flow model, $\bar{\chi}(z)$ is given either by Eq. \eqref{eq:comdistlcdm} or by Eq. \eqref{eq:comdisteds}. Note also that in Eq. \eqref{eq:qzfull}, at early times ($z\ggg 1$), $\Tilde{q}(\lambda(z))\rightarrow 1/2$, which means that the deceleration parameter measured in the tilted frame approaches its value in the CMB frame. This behaviour is expected since, as we discussed in the Introduction, the peculiar velocities and their effects fade away on large wavelengths, assuming that the universe approaches an exact FRW model on large scales. This is illustrated in Fig. \ref{fig:q_distros}, where we show the deceleration parameter $\Tilde{q}(\lambda(z))$ for both the $\Lambda$CDM (right panel) and the EdS (left fpanel) line-of-sight comoving distance as a function of the redhsift $z$, superimposed with the standard $\Lambda$CDM scenario.

\begin{figure*}
\centering
\includegraphics[width =0.7\textwidth]{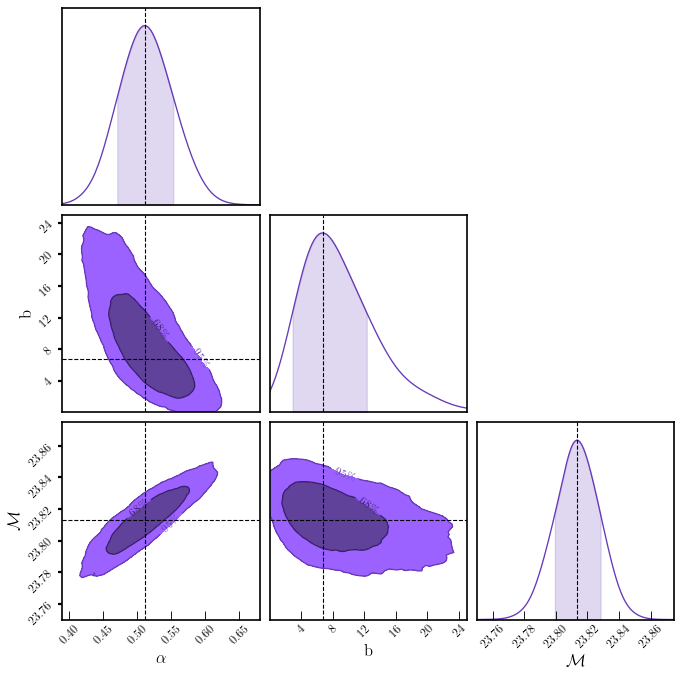}
\caption{One-dimensional and two-dimensional posterior distributions on the parameters $\alpha$, $b$ and $\cal{M}$ of the parametrization \eqref{eq:qzfull} using the EdS line-of-sight comoving distance \eqref{eq:comdisteds}. The shaded area of the histograms shows the $68\%$ error on the parameters. The contours represent the $68\%$ and $95\%$ confidence levels. The histograms have been smoothed by applying the Gaussian Kernel Density Estimation (KDE) feature from the \texttt{ChainConsumer} package. The dashed lines denote the best fit values of the parameters of the model (maximum likelihood method).}
\label{fig:contours_EdS}
\end{figure*} 

\begin{figure*}
\centering
\includegraphics[width =0.75\textwidth]{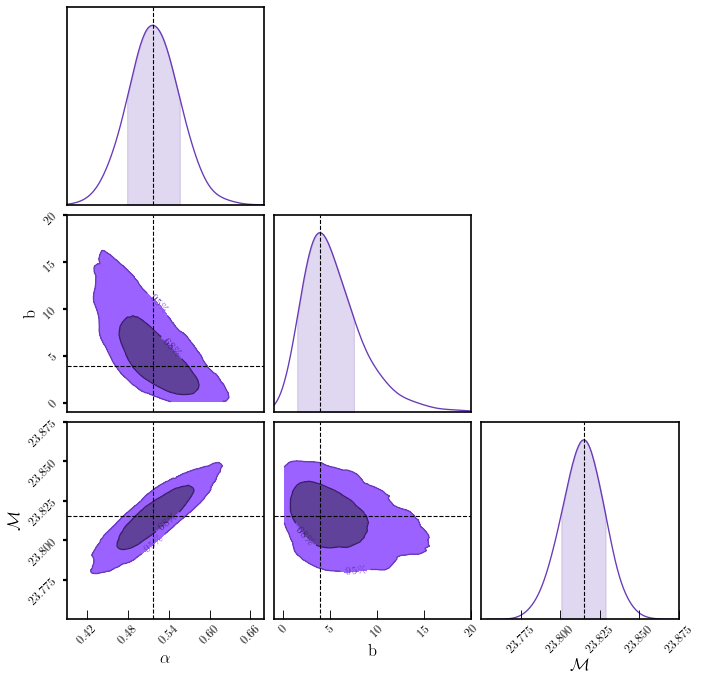}
\caption{Same as Fig. \ref{fig:contours_EdS}, but for a \lcdm line-of-sight comoving distance \eqref{eq:comdistlcdm}.} 
\label{fig:contours_LCDM}
\end{figure*} 

\subsection{Numerical Results}
Now we are ready to apply the maximum likelihood method utilizing the aforementioned equations. We construct two separate codes, one written in Python and one written in Mathematica, and apply the maximum likelihood method for the standard $\Lambda$CDM case as well as for the deceleration parameter of the tilted cosmological model assuming a $\Lambda$CDM and an EdS line-of-sight comoving distance $\bar{\chi}(z)$. In this subsection, we present only the derived results of the Python code, however the two publicly available codes are fully consistent with each other. In particular, for the standard $\Lambda$CDM scenario we obtain ${\cal{M}}=23.809 \pm 0.011$ and $\Omega_{0m}=0.299 \pm 0.022$ (see the first row of Table \ref{table:bfres}) in agreement with previous studies \cite{Pan-STARRS1:2017jku,Zhao:2019azy,Kazantzidis:2020tko,Kazantzidis:2020xta}. The maximum likelihood method for this case is applied using the corresponding expansion rate which is 
\begin{equation}
    H^2(z)=H_0^2 \left[\Omega_{0m}(1+z)^3 +(1-\Omega_{0m})\right]\,.
    \label{eq:hzlcdm}
\end{equation}
Then, we use Eq. \eqref{eq:hzlcdm} to compose the theoretically predicted apparent magnitude $m_b(z)$ through the Hubble free luminosity distance $D_L(z)$ and substitute it in the $V^i_{Panth.}$ deriving as a result the relevant $\chi^2_{\rm min}$ which is minimized in the context of the maximum likelihood method. Equivalently one can start directly from Eq. \eqref{eq:hubratqz}, substituting the form of the deceleration parameter for the standard \lcdm case which is 
\begin{equation}
    q_{\Lambda CDM}(z)=\frac{\left[\Omega_{0m}(1+z^3)-2(1-\Omega_{0m})\right]}{\left[2(\Omega_{0m} (1+z)^3+1-\Omega_{0m})\right]}\,,
\end{equation}
and solving the integral obtain the corresponding evolution of the expansion rate \eqref{eq:hzlcdm}.

In order to distinguish between each case for the tilted cosmological models, we denote the $\Tilde{q}(\lambda(z))$ parametrization assuming a $\Lambda$CDM background as Tilted-$\Lambda$ (T-$\Lambda$) while the $\Tilde{q}(\lambda(z))$ parametrization assuming an EdS background as Tilted-$EdS$ (T-$EdS$) where $\Tilde{q}(\lambda(z))$ is given by Eq. \eqref{eq:qzfull}. The results of the maximum likelihood method, \ie the best fit values and the corresponding $\chi_\textrm{min}^2$ values of the two tilted cosmological models as well as that of the $\Lambda$CDM scenario are illustrated in Table \ref{table:bfres}. In Table \ref{table:bfres}, we also include the ratio $\chi^2_{red} \equiv \chi_\textrm{min}^2/dof$, where $dof$ corresponds to the degrees of freedom and in order to achieve a good fit to the Pantheon data this ratio needs to be lower than unity.

For the determination of the best fit parameters, we minimize Eq. \eqref{eq:chi2} using the Python data fitting library \texttt{lmfit} \cite{newville_matthew_2014_11813}. Also, Fig. \ref{fig:q_distros} is produced with \texttt{matplotlib} \cite{Hunter:2007ouj}. For the construction of the posterior probability distributions of the parameters and the contours that are shown above, we apply an algorithm relying on the open-source Python package \texttt{emcee}, an implementantion of the Affine-Invariant MCMC Ensemble sampler by \cite{Foreman-Mackey:2012any}. All the plots are generated using \texttt{ChainConsumer} \cite{consumer,githubconsumer}, which analyzes the chains and produces plots of the posterior inferred from the chain distributions. For all the models of this study we use 100 random chains (walkers) and 2000 iterations (steps) for our MCMC analysis. We choose flat priors for all parameters, allowing the parameters to vary in ranges : $ 0.1<\alpha<0.9 $, $ 0<b<35 $, $ 23<{\cal{M}}<24 $. Furthermore, we construct the contours which correspond to the $1\sigma-2\sigma$ confidence levels for the T-$EdS$ parametrization in Fig. \ref{fig:contours_EdS}, while we show the same contours for the T-$\Lambda$ parametrization case in Fig. \ref{fig:contours_LCDM}.

From Table \ref{table:bfres} it is clear that, not only the two tilted cosmological models perform equally well (provide a similar $\chi^2_{\rm red}$ value) with the standard \lcdm scenario, but they also have the additional advantage of not suffering from the fine tuning problem as the standard \lcdm paradigm. Moreover, we can see that the form of the line-of-sight comoving distance $\bar{\chi}(z)$ does not affect the quality of fit to the data, since the errors of the parameter $b$ are quite large. Notice that, in the case of the tilted \lcdm cosmological model, the acceleration that the observer inside the bulk flow measures is only a local effect and happens due to the impact of peculiar motions and not due to the presence of the cosmological constant. From Fig. \ref{fig:q_distros}, it is also evident that the transition from a decelerated to an accelerated phase occurs around $z \approx 0.6$, \ie close to previous studies which assume \lcdm model \cite{SupernovaSearchTeam:2004lze, Turner:2001mx}.

This transition is determined by the parameter $b$ and is directly connected to the scale of the bulk flow. On the contrary, the parameter $\alpha$ determines the current value of the deceleration parameter as measured by an observer at the center of the bulk flow, since for $z=0$ we obtain $\Tilde{q}(\lambda(z=0)) \equiv \Tilde{q}_0=\frac{1}{2}(1-\alpha^{-1})$.  For consistency with current measurements which report $\Tilde{q}_0 \approx -0.5$ we can assume that the parameter $\alpha$ takes the generic value, $\alpha=1/2$, thus reducing the total number of parameters of the two tilted cosmological models. Applying the maximum likelihood method, we can derive the best fit and the corresponding $\chi_\textrm{min}^2$ values for this case that are very similar (\ie well within the $1\sigma$ threshold), with the derived results where the parameter $\alpha$ is free to vary. The results are also presented in Table \ref{table:bfres}.

In order to identify the optimal model, we need to take into account not only the quality of the provided fit $\chi^2_{red}$ of Table \ref{table:bfres}, but we also need to consider the number of free parameters of each model used to obtain the particular $\chi^2_{red}$ value. Even though, the best choice between the different information criteria that have been presented in the literature is not straightforward \cite{Liddle:2004nh} we use the most popular ones. The Akaike Information Criterion (AIC) \cite{AkaikeCrit,Liddle:2004nh,Nesseris:2012cq} defined as

\begin{centering}
\begin{table*}
\caption{The goodness-of-fit $\chi_\textrm{min}^2$ along with the corresponding $AIC, \, BIC$ values and the differences $\Delta AIC, \, \Delta BIC $ for the three cosmological models in question using the full Pantheon dataset.}
\label{table:AIC/BICfull}
\resizebox{0.8\textwidth}{!}{
\begin{tabular}{|c|c|c|c|c|c|c|c|}
\hline
Model & $\chi_\textrm{min}^2$ & $\chi_\textrm{red}^2$ & $AIC$ & $BIC$ & $\Delta AIC$ & $\Delta BIC$\\
\hline 
\rule{0pt}{3ex}
\textbf{$\mathbf{\Lambda}$CDM}  & $\mathbf{1026.67}$ & $\mathbf{0.981}$ & $\mathbf{1030.67}$ & $\mathbf{1040.58}$ & $\mathbf{-}$ & $\mathbf{-}$\\
T-$\Lambda$ & {$1026.69$} & {$0.982$} & {$1032.69$} & {$1047.55$} & $ {2.02}$ & ${6.97}$\\
T-$\Lambda$ ($\alpha$ fixed) & $1027.21$ & $0.982$ & $1031.21$ & $1041.12$ & $0.54$ & $0.54$\\
\textbf{T-$\mathbf{EdS}$} & $\mathbf{1026.76}$ & $\mathbf{0.982}$ & $\mathbf{1032.76}$ & $\mathbf{1047.62}$ & $\mathbf{2.09}$ & $\mathbf{7.04}$\\
\textbf{T-$\mathbf{EdS}$ ($\alpha$ fixed)} & $\mathbf{1027.05}$ & $\mathbf{0.982}$ & $\mathbf{1031.05}$ & $\mathbf{1040.96}$ & $\mathbf{0.38}$ & $\mathbf{0.38}$\\
\hline
\end{tabular}
}
\end{table*}
\end{centering}

\be 
AIC \equiv -2 \, \ln \mathcal{L}_\textrm{max}+2\, p_\textrm{tot}=\chi_\textrm{min}^2 +2\, p_\textrm{tot}  \label{eq:AIC}
\ee
where $p_\textrm{tot}$ corresponds to the total number of free parameters of the considered model and $\mathcal{L}_\textrm{max}$ corresponds to the maximum likelihood of the model under consideration. Also, we implement the Bayesian Information Criterion (BIC) which was introduced by \cite{BayesCrit,Liddle:2004nh,Nesseris:2012cq} and is defined as 
\be 
BIC \equiv -2 \, \ln \mathcal{L}_\textrm{max} + p_\textrm{tot} \, ln(N_\textrm{tot}) \label{eq:BIC}
\ee

Using the definitions \eqref{eq:AIC} and \eqref{eq:BIC} we construct the differences $\Delta AIC$ and $\Delta BIC$ of the models in question with respect to \lcdmnospace. According to the calibrated Jeffreys’ scales \cite{jeffrey}, if $0<|\Delta AIC| \leq 2$, then the confronted models can be interpreted as consistent with each other, while if $|\Delta AIC| \geq 4$ it is an indication that the model with the larger AIC value is disfavored by the data. Similarly, if $0<|\Delta BIC| \leq 2$ then the model with the larger BIC value is weakly disfavored by the data, while for $2<|\Delta BIC| \leq 6$ $\left(|\Delta BIC|>6 \right)$ the model with the larger BIC values is strongly (very strongly) disfavored. The specific differences of the studied cosmological models are shown in Table \ref{table:AIC/BICfull}.

According to the AIC, if $\alpha$ is a free parameter, the two tilted cosmological models T-$\Lambda$ and T-$  EdS$ seem to be consistent with $\Lambda$CDM. On the contrary, according to the BIC which penalizes more harshly any extra degrees of freedom, the two tilted cosmological models seem to be strongly disfavored. However, if we fix $\alpha$ to the generic value $\alpha=1/2$, the two cosmological models give $\Delta AIC$ and $\Delta BIC$ significantly lower than unity, displaying that they are equally supported
by the Pantheon sample as the standard $\Lambda$CDM model does.

\section{Discussion and conclusions}

\label{sec:concl}
We have introduced and studied two novel parametrizations of the deceleration parameter in the context of a tilted universe, with two families of relatively moving  observers. The first family are the idealised observers following the smooth universal expansion, whereas the second are the tilted observers located inside a bulk flow that moves relative to the Hubble expansion with finite peculiar velocity. Due to their relative motion, the two observers assign different values to their deceleration parameters. In fact, observers living inside locally contracting bulk flows can measure negative deceleration parameter, while the host universe is globally decelerating. Although the accelerating effect is a local artefact of the observers peculiar motion, the affected scales can be large enough to create the false impression of recent global acceleration.

The value of the locally measured deceleration parameter, as well as the scale where its sign changes from positive to negative, namely the transition length, depend on the local contraction rate ($\tilde{\theta}$) of the bulk peculiar flow. The latter, however, lies well beyond our current observational capabilities. To address the problem, we introduced a two-parameter function for the bulk-flow contraction rate (of the form $\tilde{\theta}=\tilde{\theta}(\lambda)$ -- see Eq.~(\ref{eq:volscal})), which is both mathematically simple and has sound physical motivation. Setting $\lambda \equiv \bar{\chi}(z)$, where $\bar{\chi}(z)$ is the line-of-sight comoving distance, we then obtained an expression of the form $\Tilde{q}=\Tilde{q}(\lambda(z))$ for the local deceleration parameter (see Eq.~ \eqref{eq:qzfull}), which could be directly constrained from the SnIa data.

Employing the latest publicly available SnIa compilation, namely the Pantheon sample, we successively assumed a $\Lambda$CDM and an Einstein-de-Sitter bulk-flow model and applied the maximum likelihood method for the two parametrizations. Our findings are summarized in Table \ref{table:bfres}. It is important to note that, in the case of the tilted \lcdm cosmological model, the local accelerated expansion that the bulk flow observer measures, is mainly due to the peculiar motion relative to the Hubble flow and not due to the cosmological constant. Comparing them to the standard \lcdm paradigm, we found that the three different models perform equally well (provide a similar $\chi^2_{\rm red}$ value, as indicated in the last column of Table \ref{table:bfres}) and that the form of the comoving line-of-sight distance $\bar{\chi}(z)$ does not affect significantly the derived quality of fit. However, the two tilted cosmological models have an additional parameter compared to the standard scenario. Taking into account appropriate statistical criteria such as the Akaike information criterion (AIC) and the Bayesian information criterion (BIC), we demonstrated that the three models are equally consistent according to AIC (see the sixth column of Table \ref{table:AIC/BICfull}). By contrast, BIC clearly favours the $\Lambda$CDM scenario (see the last column of Table \ref{table:AIC/BICfull}), because it has one less degree of freedom.

Nevertheless, although the $\Lambda$CDM model is still the leading cosmological paradigm, it faces a number of fundamental challenges~\cite{Perivolaropoulos:2021jda}, which are not accounted for by the aforementioned statistical criteria. Most importantly, the $\Lambda$CDM scenario does not provide a physical explanation neither for presence and the nature of the vacuum energy, nor for the fine tuning of its value. In stark contrast, tilted cosmologies, even when applied to a simple Einstein-de Sitter background, can reproduce the observed acceleration history of the universe naturally. In particular, applying the Pantheon data to our tilted Einstein-de Sitter model, recovered both the early deceleration and the late acceleration phases of the universe (see Fig. \ref{fig:q_distros}). This happened naturally, without appealing to exotic forms of matter, or introducing a cosmological constant and without any fine-tuning or coincidence problems. The tilted scenario, with its simple mathematical manifestation, can explain the recent accelerated expansion of the universe by accounting for the consequences of the bulk peculiar motions, which dominate the linear kinematics of the local Universe.

It is also worth noting that the statistical criteria that test the observational viability of a theoretical model take into account only the number of free parameters and do not take into consideration the physical motivation behind each model. Here, BIC favoured the fine-tuned $\Lambda$CDM model over the physically motivated tilted Einstein-de Sitter universe. The situation changed, however, when one of the two free parameters in Eq.~(\ref{eq:qzfull}) was fixed in advance. Setting $\alpha=1/2$ in particular ensured that  $\tilde{q}_0\simeq-1/2$  in agreement with the observations. Then, the tilted Einstein-de Sitter universe (T-$EdS$) and the standard \lcdm model achieved a similar quality of fit and are equally supported by the Pantheon data according to BIC, as it can be seen by the $\Delta BIC$ difference presented in Table \ref{table:AIC/BICfull}.

Future work, which will extend and refine this study, should include alternative parametrisation profiles for Eq. \eqref{eq:volscal}, as well as the use of different types of cosmological data. Those from the Pantheon+ dataset~\cite{Riess:2021jrx}, the SnIa compilation on~\cite{Dhawan:2021hbt}, or from SnIa and from Sn of type II~\cite{Stahl:2021mat}, should help to constrain further the parameters of the tilted model. 

From the observational view point, measuring the magnitude and the sign of the local volume scalar ($\tilde{\theta}$), which provides the spatial divergence of the bulk-flow velocity (recall that $\tilde{\theta}=\tilde{\rm D}^a\tilde{\upsilon}_a$), will directly  link both $\tilde{q}$ and $\lambda_T$ to the observations. Further insight into the bulk-flow kinematics and their implications will come from future estimates of the peculiar shear and the peculiar vorticity ($\tilde{\sigma}_{ab}$ and $\tilde{\omega}_{ab}$ respectively). In this effort, measurements of the so-called ``redshift drift'' (or ``velocity shift''), namely of temporal variations in the redshifts of distant sources~\cite{Liske:2008ph,Alves:2019hrg}, could also help.

Refining the bulk-flow data may also allow to reconstruct their internal kinematics and thus link the parameters $\alpha$ and $b$ employed in the (theoretically motivated) profile of $\tilde{q}$ (see Eq.~(\ref{eq:qzfull})) directly to the observations. Last, but not least, is the systematic search for anisotropies in the sky distribution of the deceleration parameter. Recall that a key prediction of the tilted scenario is the existence of an apparent (Doppler-like) dipolar anisotropy in the $\tilde{q}$-distribution, triggered by the observers peculiar motion.

We expect that the ongoing~\cite{Kourkchi:2020iyz, Dhawan:2021hbt} and the forthcoming~\cite{2019ApJ...873..111I} wide-field surveys should achieve exquisite precision in their bulk-flow measurements and in so doing provide us with accurate kinematic estimates of the observed peculiar velocity fields.\\

\section*{Acknowledgements}
This work was supported by the Hellenic Foundation for Research and Innovation (HFRI - Progect No: 789). The authors also acknowledge support by the IT Center of the Aristotle University of Thessaloniki (AUTh) throughout this study. We would also like to thank Eoin O Colgain for helpful comments and K.A. wishes to thank Nikolaos Karnesis for many helpful discussions concerning the MCMC analysis.

\section*{Data Availability}

The data used in the current analysis correspond to the Pantheon compilation of SnIa that are provided in  \url{https://github.com/dscolnic/Pantheon} and described in detail in Ref. \cite{Pan-STARRS1:2017jku}. The numerical files for the reproduction of the figures can be found in \href{https://github.com/lkazantzi/tilted-cosmology}{this} github repository under the MIT license.


\bibliographystyle{mnras}
\bibliography{Bibliography} 





\bsp	
\label{lastpage}
\end{document}